# Enhancing Fenton-like Photo-degradation and Electrocatalytic Oxygen Evolution Reaction (OER) in Fe-doped Copper Oxide (CuO) Catalysts

Suresh Chandra Baral[1], Dilip Sasmal[1], Sayak Datta[2], Mange Ram[3], Krishna Kanta Haldar[3], A. Mekki[4,5] and Somaditya Sen[1*]

[1]*Department of Physics, Indian Institute of Technology Indore, Indore, 453552, India*

[2]*Department of Metallurgical Engineering and Material Science, Indian Institute of Technology Bombay, Maharashtra, 400076, India*

[3]*Department of Chemistry, Central University of Punjab, Bathinda, 151401, India*

[4]*Department of Physics, King Fahd University of Petroleum and Minerals, Dhahran 31261, Saudi Arabia*

[5]*Interdisciplinary Research Center for Advanced Material, King Fahd University of Petroleum and Minerals, Dhahran 31261, Saudi Arabia*

\* Corresponding author: sens@iiti.ac.in

**Abstract:**

Although hydrogen generation by water electrolysis is the cheapest of all other available sources, water splitting still occurs with sluggish kinetics. It is a challenging barrier for $H_2$ production on a large scale. Moreover, research is still underway to understand the oxygen evolution reaction (OER) and design the catalysts with improved OER performance. Herein, we report the synthesis, characterization, and OER performance of Iron-doped copper oxide (CuO) as low-cost catalysts for water oxidation. The OER occurs at about 1.49 V versus the RHE ($\eta$ = 260 mV) with a Tafel slope of 69 mV dec$^{-1}$ in a 1 M KOH solution. The overpotential ($\eta_{10}$) of 338 mV at 10 mA cm$^{-2}$ is among the lowest compared with other copper-based materials. The catalyst can deliver a stable current density of >10 mA cm$^{-2}$ for more than 10 hours. Additionally, wastewater treatment, particularly synthetic dye wastewater, is vital for preventing water scarcity and adverse effects on human health and ecotoxicology. The as-synthesized catalysts are also utilized for Fenton-like photo-degradation under low-power visible household LED lights toward the most commonly industrially used simulated Methylene blue dye wastewater. Almost complete degradation of the MB dye has been achieved within 50 minutes of visible light irradiation with a first-order rate constant of 0.0973 min$^{-1}$. This dual functionality feature can open new pathways as a non-noble, highly efficient, and robust catalyst for OER and wastewater treatments.

**Keywords:**

1. **Introduction:**

The growing energy demand and rapid climate changes in recent times have forced scientists to develop eco-friendly solutions for clean energy applications [1], [2]. Hydrogen generated from water splitting using renewable energy sources is a clean fuel used in fuel cells, internal combustion engines, and energy storage [3]. Electrocatalytic hydrogen production using non-precious metal catalysts with renewable energy inputs (Solar or wind) is crucial for economically viable clean hydrogen production [4], [5], [6], [7]. Again, the advancement of clean hydrogen production is significantly impeded by the sluggish kinetics of the oxygen evolution reaction (OER) [8]. Metal oxides, preferably based on non-noble transition metals (such as Co, Ni, and Mn), are the most popular candidates and thus have been extensively investigated for OER with renewed interest during the last decade [9], [10], [11], [12], [13], [14]. In that context, a few recent examples have indicated that CuO-based materials could display interesting performances for water oxidation due to good electrical conductivity, nominal toxicity, and low cost [15], [16], [17], [18]. However, the performance of Cu-derived oxide catalysts for the OER is less efficient than that of Ni/Co-based catalysts, highlighting the significant improvement in the OER activity. [15], [19]. Huan *et al.* prepared a copper/copper oxide nanostructured material derived from conductive copper foam passivated by a copper oxide layer and further nanostructured by electrodeposition of CuO nanoparticles [20]. The generated electrodes were highly efficient for catalyzing selective water oxidation to dioxygen with an overpotential of 290 mV at 10 mA cm$^{-2}$ in 1 m NaOH solution. Another CuO/C Composite has been reported for efficient OER electrocatalysts in 1M KOH solution. The electrocatalyst only needed a potential of 1.49 V versus RHE (reversible hydrogen electrode) to achieve a current density of 20 mA/cm$^2$ with a Tafel slope of 115 mV dec$^{-1}$ [21]. Again, doping is an extensively explored technique that can modify the electronic band profile of transition metal oxides, resulting in the development of hybrid materials with beneficial attributes and uses [22]. Element doping can modulate the valence states through synergetic effects [19], [23], [24], [25], [26], [27], [28], [29]. Recently, Mishra *et al.* synthesized co-doped CuO microstructures using a microwave-assisted solvothermal technique. Reportedly, Co-doping markedly enhanced CuO's catalytic property for OER with an optimum 5% Co doping, the overpotential to obtain 10 mA cm$^{-2}$ is found to be reduced to 120 from 320 mV for pristine CuO, underlying the importance of doping in an inexpensive host CuO [30]. Xu *et al.* followed a cation exchange strategy to prepare a single-atom noble-metal (Rh) doped CuO nanowire arrays supported on a copper foam and reported an ultralow overpotential of 197 mV at 10 mA cm$^{-2}$ in 1 M KOH [31].

Additionally, wastewater treatment, particularly synthetic dye wastewater, is vital for preventing water scarcity and adverse effects on human health and ecotoxicology [32], [33], [34]. Along with many technologies, photocatalytic degradation of such dye-contaminated wastewater has emerged as a viable solution [35]. However, rapid, inexpensive, and sustainable catalysts for industrial-scale applications still need to be expanded. Pure and doped CuO has been utilized as visible photocatalysts for organic pollutant degradation [36], [37], [38], [39], [40], [41], [42], [43]. In a recent study, Mo-doped CuO was synthesized and utilized as a visible light photocatalyst to degrade MB dye in an aqueous solution. Reportedly, 91% degradation was achieved within 150 min of light irradiation [44]. Iron (Fe) doping in CuO seems to help increase the band gap, conductivity, and thus overall catalytic activity [45], [46]. Recently, transition metal (Zn and Fe)-doped CuO nanosheets for enhanced visible-light photocatalysis have been reported. Still, only within 270 min, Zn and Fe-doped CuO nanosheets degraded 66 % and 73 % of MB dye, respectively, while CuO nanosheets alone achieved a degradation rate of 62 %. In recent years, Fenton-like systems have been adopted to overcome the limitations of the traditional Fenton process in activating $H_2O_2$ to enhance organic pollutant mineralization [47], [48], [49], [50], [51]. Our previous study reported a fanton-like photodegradation for enhanced organic pollutants mineralization using CuO Nanosheets [52]. The effect of Fe-doping in CuO for Fenton-like photodegradation has not been investigated so far. Thus, a non-noble, highly efficient, and robust catalyst for OER and wastewater treatments is required.

Here, we report the synthesis, characterization, and application of Iron-doped copper oxide (CuO) as low-cost catalysts for OER and wastewater treatments. The OER occurs at about 1.49 V versus the RHE ($\eta$ = 260 mV) with a Tafel slope of 69 mV dec$^{-1}$ in a 1 M KOH solution. The overpotential ($\eta_{10}$) of 338 mV at 10 mA cm$^{-2}$ is among the lowest compared with other copper-based materials. The catalyst can deliver a stable current density of >10 mA cm$^{-2}$ for more than 10 hours. The as-synthesized catalysts are also utilized for Fenton-like photo-degradation under low-power visible household LED lights toward the most commonly industrially used simulated Methylene blue dye wastewater. Almost complete degradation of the MB dye has been achieved within 50 minutes of visible light irradiation with a first-order rate constant of 0.0973 min$^{-1}$. We hope this dual functionality feature can open new pathways as a non-noble, highly efficient, and robust catalyst for OER and wastewater treatments.

## 2. Materials and methods

### a. Materials and Synthesis

The sol-gel method is used to synthesize $Cu_{1-x}Fe_xO_{1\pm\delta}$ (CFO) nanopowders. The samples are named as C0 (x = 0), CF1 (x = 0.0156), and CF2 (x = 0.0312). The precursors used for the synthesis of samples were copper (II) oxide (CuO, 99.97%, Alfa Aesar) and iron nitrate nonahydrate [Fe $(NO_3)_3.9H_2O$, 99%, Alfa Aesar]. Stoichiometric ratios of CuO were dissolved in deionized water (DIW, 18.2 mega-ohm-cm) with dilute nitric acid ($HNO_3$) added. Under constant stirring, a pale blue color solution was formed at room temperature. A calculated amount of iron nitrate crystallites was added to this solution and stirred vigorously. After the homogenous mixing of the precursors, a calculated amount of citric acid and ethylene glycol were added in a 1:2 ratio to the cationic solution under vigorous stirring. The citric acid-glycerol combination acted as a gelling agent by forming polymeric chains. The homogeneous distributions of cations attach to the polymeric chains; thereby, the cations' homogeneity can be frozen. The polymers also act as fuel when burnt in the air. The resultant solution is heated on a hot plate for ~ 6 hours to obtain the gel. The gels were dehydrated and burned in the air to obtain black powders. The obtained black powders were denitrified and decarbonized by heating at 450 °C for 6 hours. The black nanopowders (nps) were grounded and were annealed at 600 °C for 2 hours. The obtained nps were ground again and are the final CFO samples. The structural and physical characterizations were carried out on CFO nps before exploring the photocatalytic dye degradation and electrocatalytic OER experiments.

### b. Structural, Vibrational, and Electronic Characterization

X-ray diffraction (XRD) analysis of the CFO nps was performed using a Bruker D2 Phaser X-ray diffractometer with Cu-$K_\alpha$ radiation ($\lambda$ = 1.54 Å) to determine the structural details. The lattice parameters, bond lengths, and bond angles were estimated by performing Rietveld refinement of the XRD data using GSAS-2 software. The phonon modes of all the samples at room temperature were obtained using a HORIBA Scientific Lab-RAM HR Evolution Raman Spectrometer. A laser light source operating at 633 nm with a power output exceeding 300 mW was used. The surface morphology of all samples was characterized using field emission scanning electron microscopy (FESEM, JEOL, JSM-7610 F). The optical bandgap was obtained using a UV-vis-NIR Shimadzu 2600 Spectrophotometer. The pore size distribution, specific surface area, and pore volume were estimated from an $N_2$ adsorption-desorption measurement [Brunauer-Emmett-Teller (BET) surface area analysis] using a Quantachrome Autosorb iQ2 BET Surface Area and Pore Volume Analyzer. The valence state of cations and the oxygen content were analyzed from the X-ray Photoelectron Spectroscopy (XPS) data obtained using a Thermo-Scientific Escalab 250 Xi instrument equipped with a monochromatic Al $K_\alpha$ X-ray source (hv = 1486.6 eV) operating at a power of 150 W and under UHV conditions in the range of ∼$10^{-9}$ mbar. All spectra were recorded in hybrid mode, using electrostatic and magnetic lenses and an aperture slot of 300 µm × 700 µm. The wide and high-resolution spectra were acquired at fixed analyzer pass energies of 80 and 20 eV, respectively. A flood gun was used to compensate for charging effects. All electrochemical measurements were performed with Metrohm Autolab (M204 multichannel potentiostat galvanostat) using Nova 2.1.4 software.

### c. Photocatalytic Properties characterization

The photocatalysis was performed on commercially procured MB ($C_{16}H_{18}ClN_3S$, 99.9% purity, SRL chemicals, India). Hydrogen peroxide solution (30% w/v) ($H_2O_2$, 99.9% purity, Rankem, India) was procured as an oxidizing agent.

The photocatalytic activity of all the samples was examined under the visible light illumination from a household LED, emitting monochromatic light having maximum intensity at 455 nm, light intensity 100 lx ≈ 0.79 Wm$^{-2}$, which is 1266 times smaller than AM 1.5G Sun). A 10-ppm strength stock solution of pollutant methyl blue (MB) dye was prepared by mixing 10 mg of the MB dye in 1 L DIW. Five different concentrations (0.5 mg/mL, 0.75 mg/mL, 1 mg/mL, 1.25 mg/mL, and 1.5 mg/mL) of the catalyst solution were created by adding an appropriate amount (10 mg, 15 mg, 20 mg, 25 mg, and 30 mg, respectively) of CFO nps to 20 ml of MB dye solution in separate beakers. Initially, 0.115 ml (50 mM) of hydrogen peroxide ($H_2O_2$) (30% w/v) was added to this solution to enhance the photocatalytic reaction. The obtained mixed solutions were stirred continuously in the dark for 60 minutes to achieve good dispersion and adsorption equilibrium. The concentration of the remaining dye was estimated from the time-dependent absorption spectra (10-min interval) recorded using a Research India UV-visible spectrophotometer. After each step of the photocatalytic degradation of samples, a 3 mL solution was extracted from the photoreacted solution. The photoreacted solutions were centrifuged properly to extract the catalyst nps. The reduction of the intensity of the absorption spectra with time was studied for each sample. The photocatalytic degradation efficiency of the organic dyes was calculated using the following formula: $Degradation\ (\%) = [(A_0 - A_t)/A_0] * 100$ ; where $A_0$ is the absorbance of dye solution before the photo-irradiation and $A_t$ is the absorbance of solutions after photo-irradiation for a specific time t.

### d. Electrochemical measurements

A homogenous electrocatalytic ink was prepared by dispersing and sonicating 3 mg of CFO nps in a mixture of 400 μL of ethanol and 680 μL DIW. Again, 20 μL of a 5% Nafion solution was added to the ink, followed by sonication for another 20 min. Finally, 5 μL of as-prepared ink was drop-casted over a glassy carbon electrode (GCE) of 3 mm diameter and dried in a vacuum desiccator at ambient conditions. This was further used as the working electrode (WE) to study the electrocatalytic OER. The typical mass loading is 0.19 mg/cm$^2$. All the electrochemical measurements were taken at room temperature in ambient conditions using Metrohm Autolab (Multichannel-204), a standard three-electrode system equipped with Nova 2.1.4 software. A 1M KOH electrolyte solution was prepared in DI water. This solution was degassed with nitrogen gas for 30 min to remove the dissolved oxygen and other impurities. This solution was subjected to OER measurements. OER was performed in a conventional three-electrode system consisting of Hg/HgO as reference electrode (RE), platinum electrode as counter electrode (CE), and GCE as WE. Linear sweep voltammetry (LSV) and cyclic voltammetry (CV) measurements were performed in the optimized potential range of 0–1 V between the WE and RE at a scan rate of 5 mV/s to study the OER. The electrochemical impedance spectroscopy (EIS) measurements were performed at an applied potential in the frequency range of ~ 10$^5$ to 0.1 Hz. Then, the electrochemical double layer capacitance ($C_{dl}$) was measured and calculated after linear fitting in the non-Faradaic region corresponding to the cyclic voltammetry curve at different sweep rates (20, 40, 50, 60, 80, and 100 mV/s). The long-term stability of CF1 for 10 h is tested using chronopotentiometry at a constant current density of 10 mA/cm$^2$ for OER. The stability of the sample for OER activity is analyzed by comparing the change in overpotential after 2500 LSV cycles at a constant scan rate of 50 mV/s. In hydrogen production processes, the hydrogen electrode is one at which the hydrogen is produced and is generally similar to the RE. Such an electrode is called the Reversible Hydrogen Electrode (RHE). All potentials were calibrated with respect to $E_{RHE}$, which is the potential of the RHE, given by the Nernst equation: $E_{RHE} = E_{Hg/HgO} + E^{\theta}_{Hg/HgO} + 0.059*pH$, where $E^{\theta}_{Hg/HgO}$ is the standard reduction potential of Hg/HgO (0.097 V) [53], and $E_{Hg/HgO}$ is the applied potential in RE. The pH of the solution was 13.8 during the entire measurement.

$C_s$ is the specific capacitance of the flat working electrode (i.e., 40 µF/cm$^2$) [54]. This is an important parameter, as the ratio of $C_{dl}/C_s$ provides the effective electrochemically active surface area (ECSA). The anodic current density ($J_a$) and cathodic current density ($J_c$) depend on the scan rate. The rate of change of these current densities with the scan rate estimates the capacitance of the double layer, $C_{dl}$. Hence, $C_{dl}$ was determined from the slope of $J_a$ or $J_c$ vs. scan rate and then averaged.

### 3. Results and Discussion

The XRD pattern of all the samples revealed a single-phase monoclinic tenorite structure with a C2/c space group [Figure 1]. Various diffraction peaks appeared at 2θ = 32.5°, 35.5°, 38.7°, 48.8°, 53.3°, and 58.2° belonging to the (110), (002)/(11-1), (200)/(111), (20-2), (020), and (202) planes of the tenorite CuO structure [JCPDS card # 48–1548]. With Fe incorporation, there were structural modifications of the tenorite structure, as revealed by the Rietveld refinement of the XRD data using GSAS-2 software.

In tenorite (CuO), the Cu atom is in a distorted octahedral crystal field surrounded by six oxygen atoms. Two of the Cu-O bonds are larger (~2.766 Å) than the rest four (~1.948 Å). Note that the diagonally opposite oxygens are linearly connected with the centrally located copper atom. However, the three sets of such linearly connected oxygen atoms are not perpendicular to each other and are distorted with angles between the different Cu-O bonds varying from 73.23°, 89.18°, and 84.37°.

Hence, four O-atoms bonds with a Cu atom. Therefore, the Cu is most likely four-coordinated in the CuO lattice [55]. The $Fe^{3+}$ (IV) has a much lesser radius (0.63 Å) in comparison with that of the $Cu^{2+}$ (IV) (~0.71 Å). Hence, one can expect a shrinking of the lattice near the dopant $Fe^{3+}$ atom due to the smaller size of the host $Cu^{2+}$ atom. However, the charge of $Fe^{3+}$ is greater than the $Cu^{2+}$ ion and is supposed to attract a more significant amount of negatively charged $O^{2-}$ions to the lattice. It has been reported that p-type conductivity in pure CuO can be traced back to the intrinsic Cu-vacancies in the system [56]. $Fe^{3+}$ incorporation with extra charge can further lose $Cu^{2+}$ ions [57], decreasing lattice volume from 79.986 Å (C0) to 79.9719 $Å^3$ (CF1) but further adding. $Fe^{3+}$ increased the cell volume to 79.981 $Å^3$ (CF2) by pulling more oxygen into the lattice's interstitial positions. The same effect is also observed in the lattice parameter as well. The lattice parameter 'a' decreased from C0 4.46162 Å to CF1 4.66133 Å and then increased to 4.66151 Å. However, the lattice parameter 'b' increased from 3.40606 Å (C0) to 3.440651 Å (CF1) and then decreased to 3.40636 Å (CF2). The lattice parameter 'c' decreased from 5.10544 Å (C0) to 5.10429 Å (CF1) and then increased to 5.10512 Å (CF2). These variations in the lattice parameters and cell volume indicate the incorporation of the $Fe^{3+}$ ions in the CuO lattice.

The crystallite size (D) for all the samples was determined from the Scherer equation (D = 0.9 λ/FWHM*cosθ). FWHM is the full width at the half maximum intensity of the XRD peak, and θ is Bragg's diffraction angle. The D value was obtained considering the plane (110) of CuO, which was 86.73 nm for C0, then decreased to 86.52 nm after doping with 1.56% Fe (CF1), which again increased to 86.86 with the incorporation of 3.12% Fe (CF2) in the CuO system.

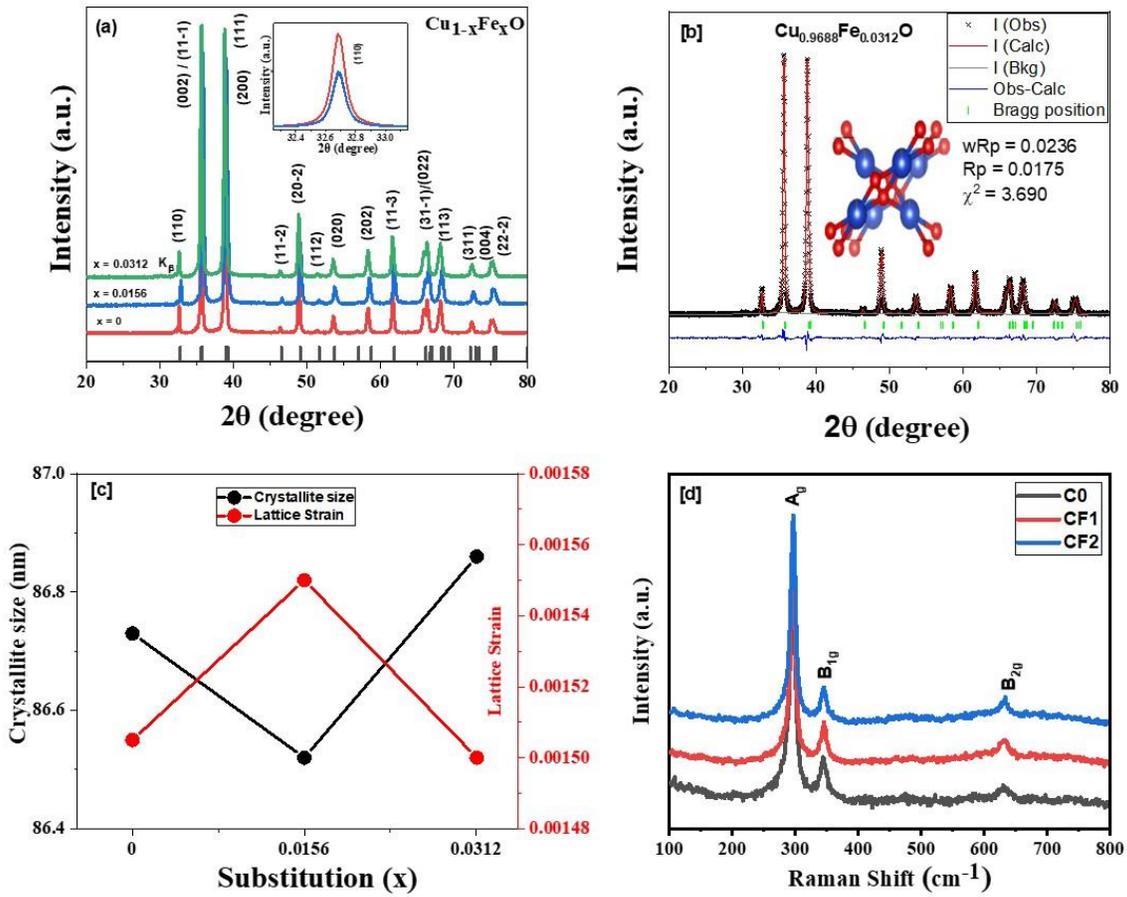

**Figure 1:** *Represents structural characterization of all the samples; (a) shows XRD pattern, while the inset shows the magnified image of the 002 planes, indicating peak shifting; (b) refined structure for the CF2 sample with the goodness of fit; (c) showing the variation of crystallite size with lattice strain; and (d) represents the Raman spectra for all the samples.*

The group theory analysis, supported by experimental results, confirms the presence of twelve vibrational modes in monoclinic CuO. These include three acoustic modes ($A_u + 2B_u$), six infrared active modes ($3A_u + 3B_u$), and three Raman active modes ($A_g + 2B$) [58]. The active Raman modes are mainly due to vibrations of oxygen atoms because of the symmetry of the crystal structure [59]. The absence of impurity peaks in the recorded Raman spectra confirms the formation of single-phase monoclinic CuO, consistent with our XRD findings. The recorded

Raman spectra revealed Raman vibration at 295.8 cm$^{-1}$ (A$_g$), 344.5 cm$^{-1}$ (B$_{1g}$), and 630.1 cm$^{-1}$ (B$_{2g}$), among which A$_g$ mode is the strongest feature for powder samples. A significant blue shift (296.6 cm$^{-1}$) is observed in the CF1 sample. However, a negligible shift (296.5 cm$^{-1}$) of the A$_g$ mode is observed in CF2. The blueshift in Raman mode (A$_g$) may indicate an increase in bond strength, which can be further related to the higher charge in the dopant Fe$^{3+}$. The higher charge of Fe$^{3+}$ in the Cu$^{2+}$ host increases the adequate bond strength due to the stronger overlapping of the electron clouds that may lead to the blue shift. The full width at half maxima (FWHM) of the A$_g$ mode decreased for CF1. A decrease in FWHM often suggests increased molecular order or crystallinity in a material [60].

FTIR spectra of the synthesized nanostructures were obtained in the wavenumber range of 400–4000 cm$^{-1}$ to analyze the surface chemistry and vibrational characteristics of the atoms [Figure 2(a)]. Literature reports standard Cu-O stretching modes characteristic of a CuO structure at 473 cm$^{-1}$ (A$_u$ mode along [1 0 1] direction), 538 cm$^{-1}$, and 595 cm$^{-1}$ (B$_u$ mode along [-101] direction) [55], [61]. The impact of Fe doping on the modes was evident. The A$_u$ mode was observed at 474 cm$^{-1}$ in C0, which blueshifts to 476 cm$^{-1}$ in CF1 and slightly redshifts to 475 cm$^{-1}$ in CF2. Shifts in the FTIR modes indicate changes in the phonon energy. A blue shift, i.e., shifting of the Raman peaks to higher frequencies, can suggest an increase in bond strength or a change in hybridization states, generally a more vital interaction within the electron clouds, leading to higher conductivity in Fe-modified CuO [57]. No peak related to the Fe-based compound has been observed, which concludes that the prepared Cu$_{1-x}$Fe$_x$O$_{1\pm d}$ has a single-phase monoclinic structure.

The UV-Vis Diffuse Reflectance Spectra for all the samples were acquired over the range of 200~800 nm. The band gap is determined from these spectra using the Kubelka-Munk function: F(R) = (1-R)$^2$/2*R [62], where R is the diffuse reflectivity of the samples. The absorption coefficient, α, is close to the F(R). The optical band gap was calculated from energy-dependent optical absorbance data using the Tauc method [63]. The band gap energy was determined by the Tauc equation: - $(\alpha h\nu)^{1/n} = A(h\nu-E_g)$ [64], where α is the absorption coefficient, ν is the photon's frequency, h is Planck's constant, A is a proportionality constant, and $E_g$ is the bandgap. The nature of the transition between the conduction band (CB) and valence band (VB) is determined by the power n; n = 1/2, 3/2, 2, and 3 for direct allowed, direct forbidden, indirect allowed, and indirect forbidden transitions, respectively. Doping introduces elements with different valence states and ionic sizes, leading to lattice distortions that can create defect states, which can modify the bandgap and introduce states inside the gap [65]. The positioning of these defect states and the alteration of the band edges can result in tailing near the band edge. The energy corresponding to these states is referred to as the Urbach energy ($E_U$), which can be calculated from the tail regions of the spectrum. [65]. The value of α increases exponentially with energy E = hν and is influenced by $E_U$: $\alpha = \alpha_o \cdot \exp(h\nu/E_U)$, where $\alpha_o$ is a constant. Natural logarithm on both sides results in a linear relation between ln(α) and E: $\ln[\alpha] = \ln[\alpha_o] + (E/E_U)$. This enables the estimation of $E_U$ from the linear fitting of the linear region in the ln(α) vs E plots.

The inter-atomic hybridization between Cu 3d and O 2p states results in the formation of valence and conduction bands in CuO. While the conduction band mainly comprises the Cu 3d state, the O 2p state is predominant in the valance band [66]. The band gap analysis, conducted using the Tauc method, revealed that the band gap energy continuously increased from C0 to CF2 [Figure 2b]. The $Fe^{3+}$ ion incorporation in CuO increases the optical band gap from 1.435 eV (C0) to 1.5 eV (CF1) and 1.63 eV (CF2). Again, the Urbach energy, which is associated with the defects, has decreased from 93 meV (C0) to 80 meV (CF1) but increased to 297 meV (CF2). This confirms the lowest defects in CF1 compared to C0 but the highest for the CF2 sample. This observation can be explained in two ways.

One of the reasons for this is the Burnstein-Moss, which is a doping effect frequently mentioned in literature [60], [67], [68]. This effect is observed in doped samples when hybridized states in between the dopant atoms and O are formed in the conduction band but very close to the conduction band minima. Electrons can be excited directly to these trap states, and an effective bandgap appears to increase. Initially, $Fe^{3+}$ ions can take the void of a copper vacancy ($V_{Cu}$). It has been observed that $V_{Cu}$ states are located very near the conduction band edge. Therefore, eradicating these states may imply a reduction in the possibility of a transition of an electron from the VB to these states, ultimately removing the deep trap levels and increasing the absorption coefficients for the CF1 sample. However, a further increase in $Fe^{3+}$ can lead to other defects like oxygen at the interstitial positions, as observed from the rise in Urbach energy for the CF2 sample. Several different reports also suggest Fe incorporation can remove deep trap levels and increase the absorption coefficient of CuO [60], [69]. The highest absorption coefficient in the CF1 and lowest in the CF2 sample thus can be correlated to the above possibilities.

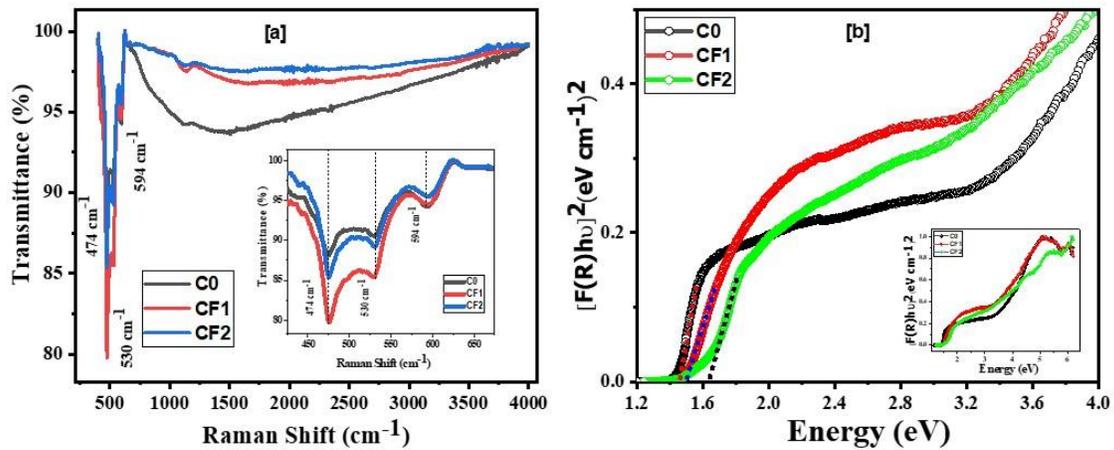

**Figure 2:** *Represents (a) the FTIR spectra, while the inset shows the magnified image in a smaller range, indicating peak shifting; (b) variation in the band gap for all the samples, where the inset shows the broad spectrum indicating the highest absorption in the case of the CF1 sample.*

The surface morphology for all the samples was obtained with the help of Field Emission Scanning Electron Microscopy (FESEM). The microscopic images of the samples obtained from FESEM are illustrated in Figure 3. All the samples show agglomerated particles ranging in size to less than a micrometer, and EDX mapping shows a homogeneous distribution of all the elements through the samples.

The Brunauer-Emmett-Teller (BET) surface areas for all the samples were analyzed using $N_2$ adsorption and desorption data [Figure 4(a)]. The specific surface area first increases rapidly from 1.49 m$^2$ g$^{-1}$ for C0 to 6.448 m$^2$ g$^{-1}$ for CF1 and then decreases to 4.891 m$^2$ g$^{-1}$ for CF2 samples. The maximum pore volume for the CF1 sample is 0.031 cm$^3$ g$^{-1}$, while the minimum pore volume for the C0 sample is 0.012 cm$^3$ g$^{-1}$. The values of the specific surface area, the average pore diameter, and the total pore volume of all the samples are given in Table 1. The results obtained from the BET study show that the surface area and pore volume for all Fe-doped samples are more significant than those of pure CuO samples. The CF1 sample has the highest surface area and pore volume among all the samples. C0 and CF2 samples have less surface area than the CF1 sample.

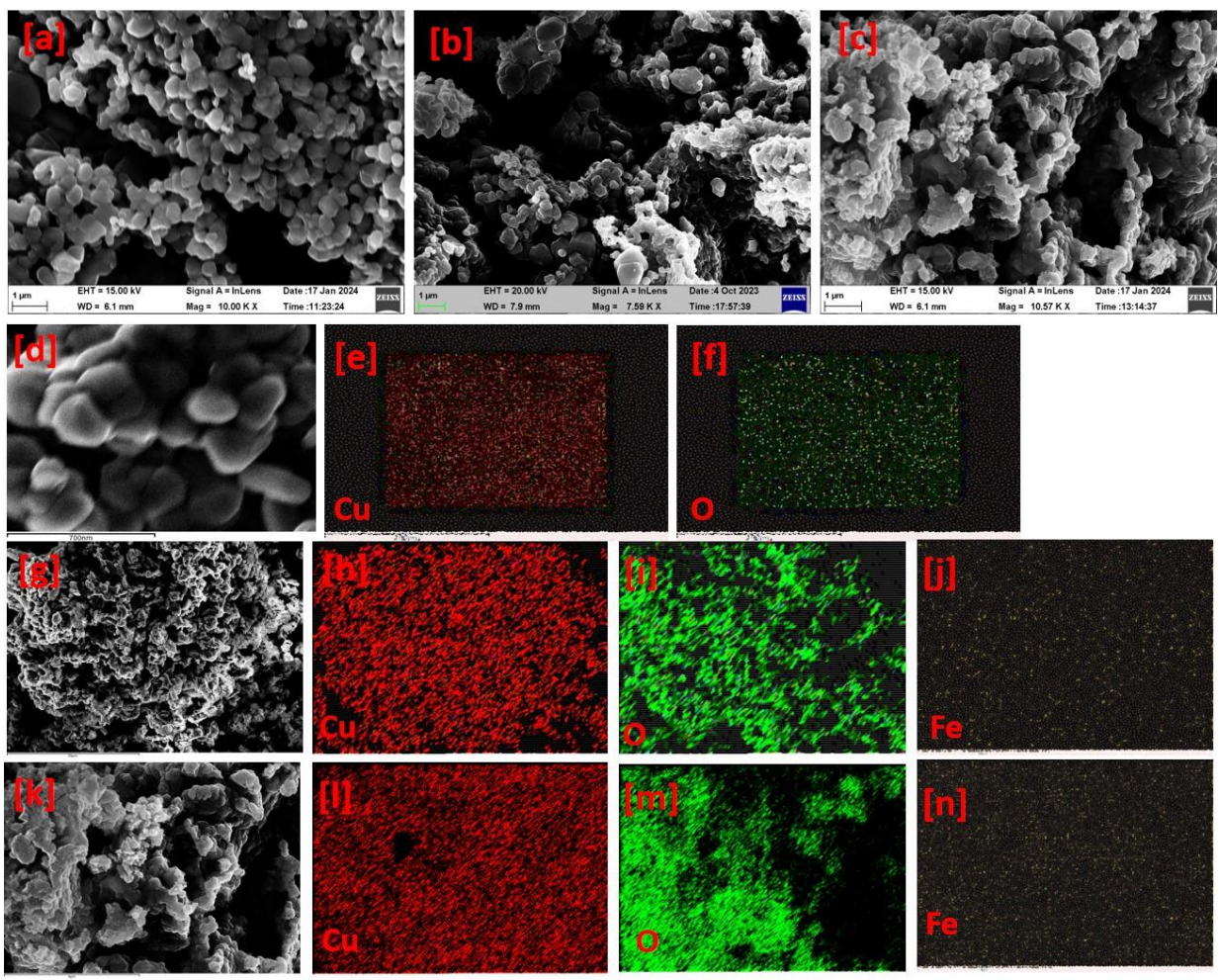

**Figure 3:** *(a, b, c) Represents the FESEM images (d, e, f), (g, h, i, j), and (k, l, m, n) showing elemental mapping for C0, CF1, and CF2 samples, respectively.*

**Table 1:** *Representation of specific surface area, average pore diameter, and total pore volume for all the samples*

| Catalyst | Specific Surface Area ($S_{BET}$) (m² g⁻¹) | Total Pore Volume ($V_{pore}$) (cm³ g⁻¹) | Average Pore Diameter ($D_{pore}$) (nm) |
|---|---|---|---|
| C0 | 1.49 | 0.012 | 3.813 |
| CF1 | 6.448 | 0.031 | 3.407 |
| CF2 | 4.891 | 0.043 | 4.302 |

The XPS technique was used to analyze pure and doped CuO's electronic state and chemical composition. The overall spectra of pure and Fe-doped CuO nanostructures are shown in Figure 4, showing only the presence of these elements. Figure (b) shows no impurity except carbon contamination from the atmosphere located at a binding energy (BE) of 284.8 eV, and Figure (d, e, f) represents the high-resolution Cu 2p core level spectra of all the samples. It can be observed that the Cu 2p core level splits into the usual spin-orbit doublet, Cu $2p_{3/2}$ and Cu $2p_{1/2}$, along with two satellite peaks, and these satellite peaks are seen at higher BE (~ 942 eV and ~ 963 eV) than the main peak, which is due to Cu $3d^9$ states [70]. In the case of the pure (C0) sample (pure CuO), the main core level peaks are observed at a BE of ~ 934.0 eV and ~ 954.0 eV with an energy separation of 20 eV. The peaks in the Fe-doped CuO sample are observed at a BE of ~ 934.1 eV and ~ 954 eV for both CF1 and CF2, corresponding to Cu $2p_{3/2}$ and Cu $2p_{1/2}$, respectively. The main peaks and the satellite are due to Cu ions in $Cu^{2+}$ oxidation in both CF1 and CF2 samples, which are in agreement with the analysis of the pure CuO sample. Figure 4(g, h, i) shows the O 1s core level spectra for all samples. For the C0 sample, the high-resolution peak was fitted with three contributions. One assigned to the lattice oxygen ($O_i$) centered at ~ 530 eV. The other two peaks were at B.E. of ~ 531 eV and ~ 533 eV, assigned to hydroxyl group oxygen/oxygen vacancy and to adsorbed water molecules, respectively. Similar fitting results of the O 1s spectra were found for the CF1 and CF2 samples. Similarly, Figure 4 (c) shows the high-resolution spectra for Fe 2p orbitals. The BE of Fe $2p_{3/2}$ and Fe $2p_{1/2}$ are about at ~ 711.8 eV and ~ 724.6 eV, respectively, indicating the presence of $Fe^{3+}$ in the catalysts [71], [72], [73]. However, quantitative information about the oxidation states of Fe could not be extracted due to the auger lines of Cu overlapping with the Fe 2p signal. In addition to the above peaks, a hydroxide peak was observed in the high-resolution Cu 2p spectra centered at ~ 936 eV. Copper hydroxide (Cu(OH)$_2$) is considered to be a corrosion product layer [74]. The area percentage of Cu(OH)$_2$ compared to the total area under Cu 2p$_{3/2}$ was 0.486 for C0 and was found to be reduced to 0.44 for CF1 and increased to 0.5 for CF2. It has also been reported that CuO is catalytically more active than Cu(OH)$_2$ [75]. On the other hand, the hydroxyl group oxygen/oxygen vacancy increases from 0.5 for C0 to 0.65 for CF1 and then decreases to 0.56 for CF2. When oxygen vacancies are crucial in enhancing the electrocatalytic activity of the oxygen evolution reaction (OER), CF1 can play an essential role as an OER electrocatalyst [76], [77].

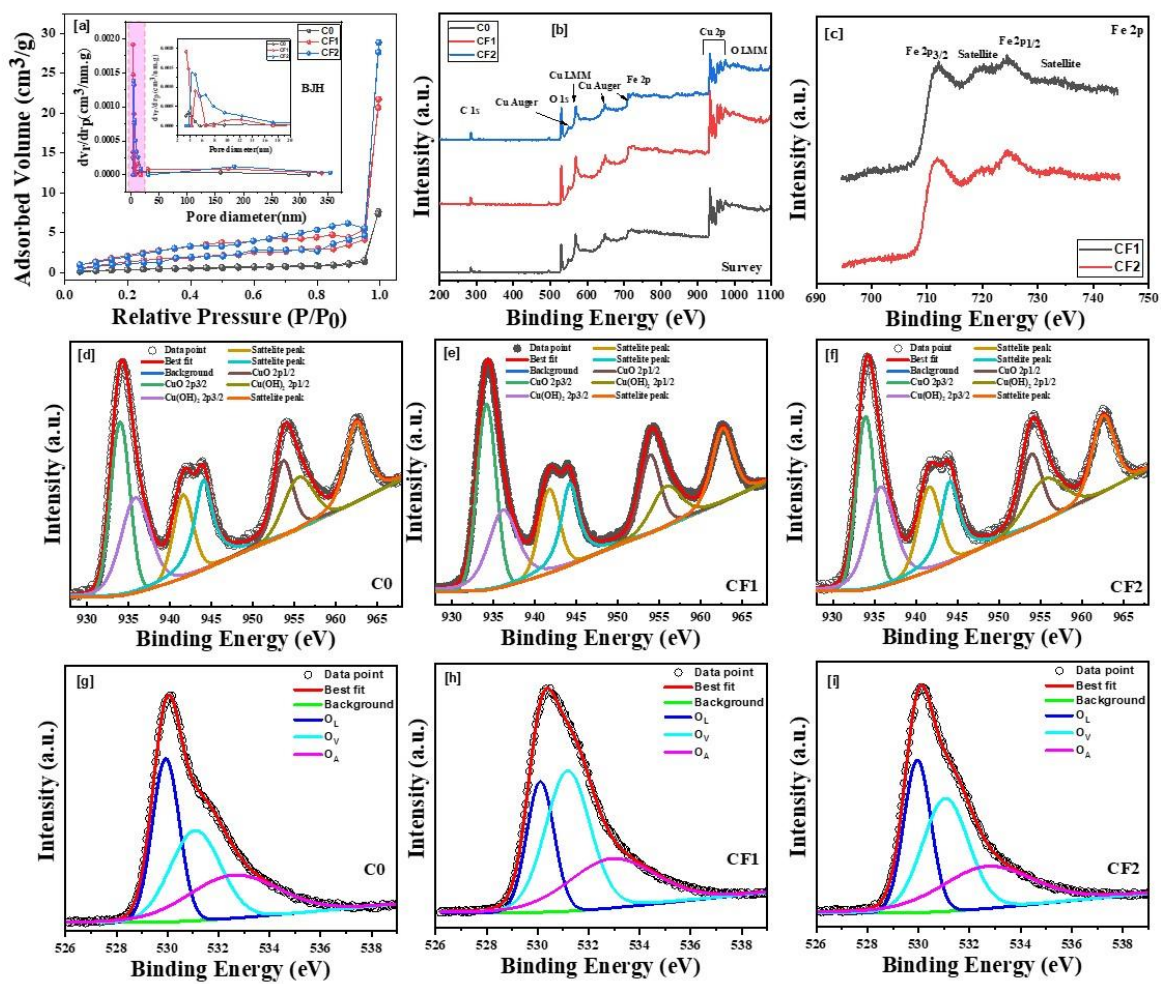

**Figure 4:** *Represents (a) the $N_2$ adsorption-desorption isotherm, where the inset shows the corresponding BJH pore size distribution, (b) shows the survey spectra, and (c) high-resolution XPS spectra of iron (Fe), (d, e, f) of copper (Cu), and (h, i) of oxygen (O) for C0, CF1, and CF2 samples, respectively.*

Dynamic light scattering (DLS) is a powerful tool for understanding particle size and dynamics, providing crucial insights into the behavior of colloidal systems. The behaviors of the samples in a colloidal solution can be characterized by three essential parameters, i.e., the Zeta potential ($V\xi$) ($\xi$-potential), the hydrodynamic size ($R_h$) [HDS], and the Polydispersity Index (PDI) which describe the Brownian motion of macromolecules in solution were extracted. $R_h$ is the effective spherical size of a molecule of the particles, which provides the same friction in the fluid as the molecule itself. The $V_\xi$ is an estimate of the surface charge of crystals, while the PDI measures the broadness of the size distribution of the particles in a sample. $R_h$ showed a random variation. The $R_h$ of C0 is $454 \pm 6$ nm, CF1 is $597 \pm 4$ nm, and CF2 is $528 \pm 3$ nm. The C0 sample revealed a PDI value of 0.3. The PDI value of CF1 was 0.394, while that of CF2 was 0.419, conferring a mid-range polydispersity nature of all the samples. Similarly, the Zeta potential ($V_\xi$) for C0 was ~-14.63 mV, whereas for CF1, it was ~-8 mV, and for CF2, it was ~-14.43 mV. The positive Zeta potential for all the samples indicates that all the samples will show better adsorption kinetics for cationic dyes.

4. **Applications**
    a. **Photocatalytic dye degradation**

The most common practice for checking photocatalytic activity is using a costly apparatus with a high-power (250~300W) UV/Xenon lamp. In the present work, an attempt has been made to test the photocatalytic activity of the nps for simulated wastewater in the presence of standard, inexpensive, commercially available household LED bulbs. The photocatalytic activity was carried out for pure and Fe-modified CuO using simulated MB dye wastewater.

All the photocatalytic experiments were carried out at room temperature (30 $^0$C) and normal atmospheric pressure [Figure 5]. Hydrogen peroxide can enhance the charge separation process by accepting an electron from the conduction band and producing OH$^-$ radicals. When irradiated under visible household LED for 50 mins, the photocatalytic degradation by pure CuO was found to be 87.7%. However, with the introduction of Fe in CuO systems, there is an increase in photocatalytic degradation up to 98.8% in the case of CF1 and 99.1% in the case of CF2, keeping the same exposure dose of LED light as before. [Figure 5(f)]. Therefore, it was found that all the Fe-doped CuO samples showed better dye degradation properties than pure CuO.

After that, the concentration of the CF2 nps was varied to see the photocatalytic degradation effect. The photocatalytic degradation was checked simultaneously for different concentrations: 0.5 mg/mL, 0.75 mg/mL, 1 mg/mL, 1.25 mg/mL, and 1.25 mg/mL. All the solutions containing the simulated wastewater and the required concentration of catalysts were subjected to 50 minutes of light irradiation [Figure 5(g)]. As the catalyst loading increases from 0.5 mg/mL to 1 mg/mL, the degradation of MB increases from 97% to 99%, then decreases to 97.5% for a catalyst with 1.25 mg/mL. The increasing dye photodegradation with increasing catalyst amount is a feature of heterogeneous photocatalysis as the number of collisions between reactants increases [78]. As the number of active sites on the photocatalyst surface increases, the formation of OH• radicals increases, which can take part in the actual discoloration of the dye solution. Beyond a certain amount of catalyst, the solution becomes turbid and thus blocks the radiation for the reaction to proceed. Therefore, the percentage degradation starts decreasing [79]. The optimum concentration has thus been taken to be 1 mg/mL.

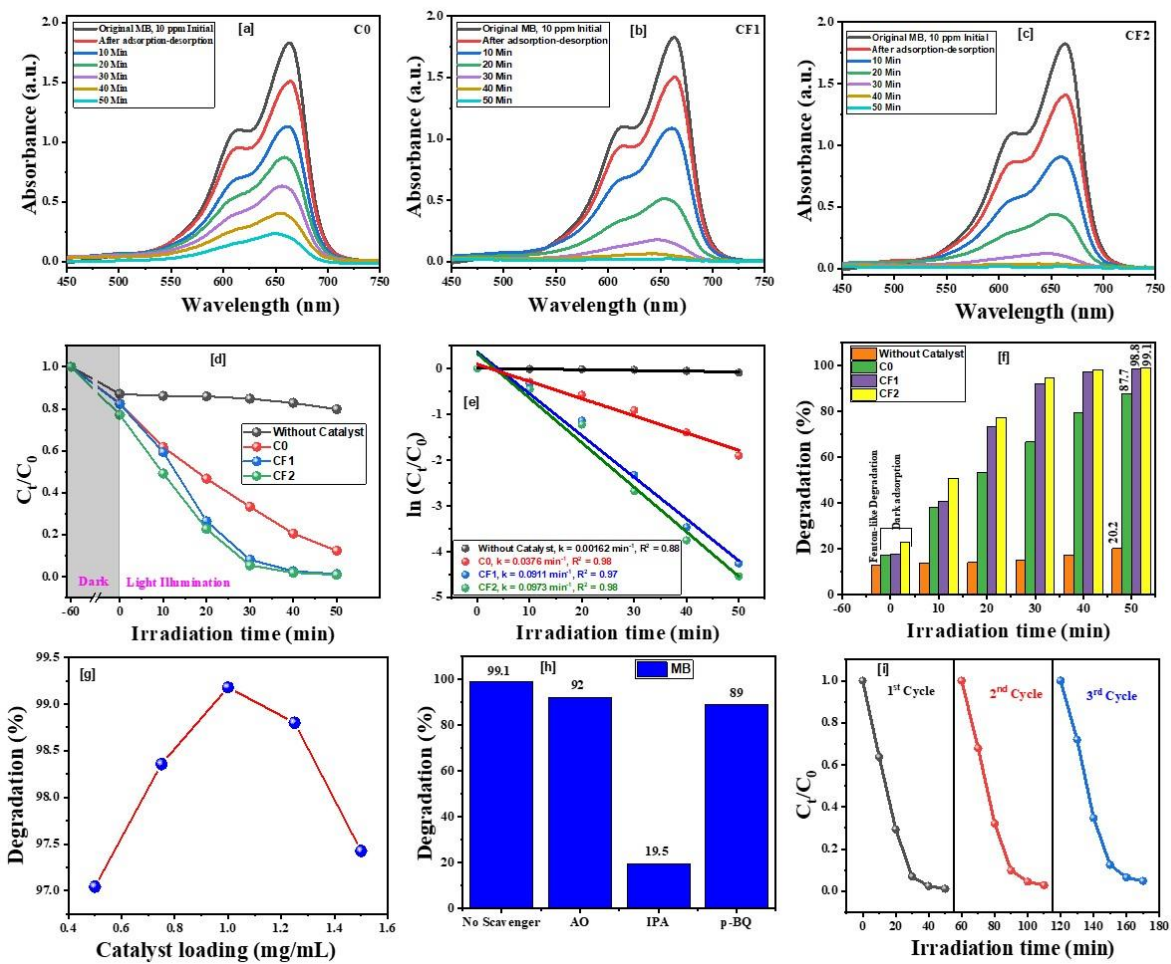

*Figure 5: Absorbance of MB solution over time for C0, CF1, and CF2, respectively, is represented by (a, b, c), and the relative concentration and calculated rate constants are represented by (d, e). The degradation percentage of all the samples is shown by (f) and (g, h, i), showing the degradation percentage for different catalyst concentrations, scavengers test, and recyclability test for the CF2 sample, respectively.*

### i. Mechanism

The detailed mechanism of dye degradation has already been explained in detail in the literature [52]. The role of $Fe^{3+}$ in CuO for catalytic degradation can be explained by the oxidation and reduction capability of the material [Figure (6)]. In this regard, the position of the valence band ($E_{VB}$) and the conduction band ($E_{CB}$) are important parameters. The values of $E_{VB}$ and $E_{CB}$ for different samples were calculated and summarized in Table 2 [55]. It can be observed that the incorporation of $Fe^{3+}$ into the copper oxide system increases its band gap, thereby altering the corresponding valence band edge and the conduction band edge. Since the standard electrode potential for $OH^-/OH^.$ is around 1.9 eV, the hole, $h^+$, from the valence band, combines with $OH^-$ to create $OH^.$ radicals, which is more thermodynamically favorable. Similarly, $OH^./H_2O_2$ has a redox potential of 0.8 eV. As a result, $OH^-$ radicals are produced when the $e^-$ from the conduction band reacts with $H_2O_2$. These radicals can then participate in the reactions. The organic dye molecule is now converted by reactive $OH^-$ radical into less hazardous by-products such as $CO_2$ and $H_2O$ [55].

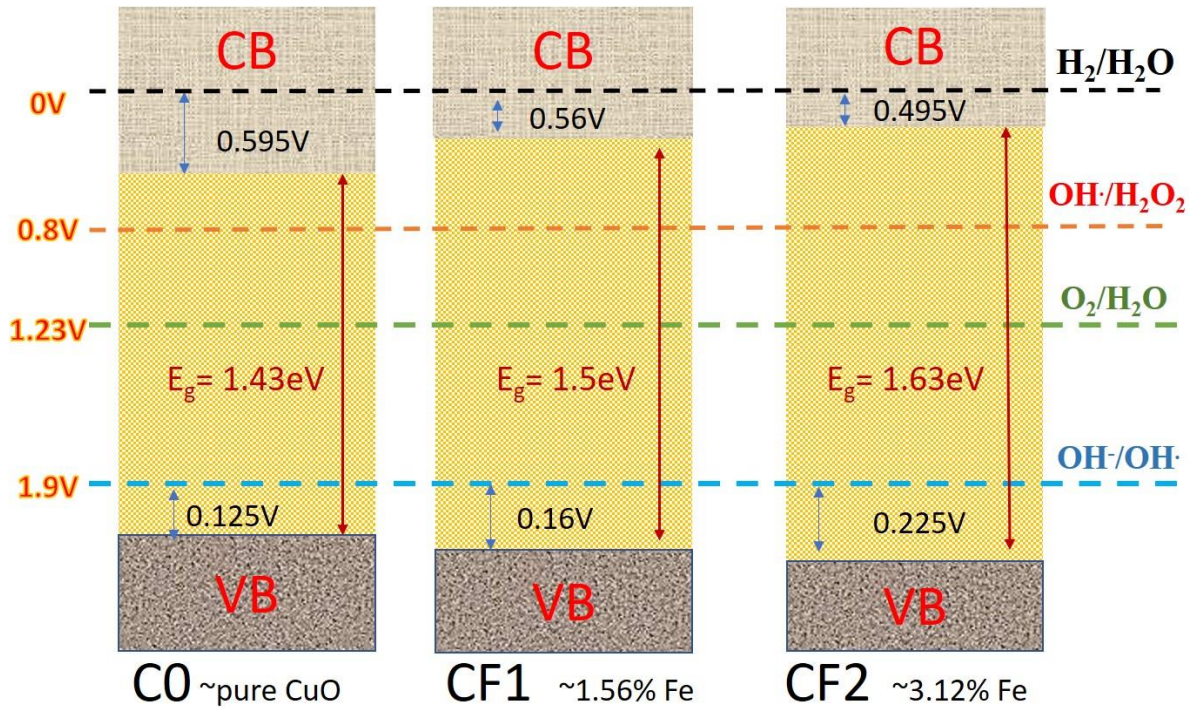

*Figure 6. Schematic representation for degradation mechanism of $Fe^{3+}$-modified copper oxide. Table 2 represents the standard $E_{VB}$ and $E_{CB}$ values for different samples.*

| Sample | $E_{VB}$ | $E_{CB}$ |
|--------|----------|----------|
| C0     | 2.025    | 0.595    |
| CF1    | 2.06     | 0.56     |
| CF2    | 2.125    | 0.495    |

Additionally, the reaction mechanism of the fast degradation of MB with the support of hydrogen peroxide and Fe-doped CuO without using any external source of irradiation with photon proceeds according to the following [80], [81], [82]. Hence, introducing Fe in CuO could further increase the Fenton-like degradation.

Step 1, $\quad Cu^{2+} + H_2O_2 \rightarrow Cu^+ + •OOH + H^+$

$Cu^+ + H_2O_2 \rightarrow Cu^{2+} + •OH + HO^-$

$Fe^{3+} + H_2O_2 \rightarrow Fe^{2+} + •OOH + H^+$

$Fe^{2+} + H_2O_2 \rightarrow Fe^{3+} + •OH + HO^-$

$Fe^{2+} + •OH \rightarrow Fe^{3+} + HO^-$

$Fe^{3+} + •OOH \rightarrow Fe^{2+} + H^+ + O_2$

$Fe^{2+} + •OOH \rightarrow Fe^{3+} + HO_2^-$

Step 2, $\quad •OOH + H_2O_2 \rightarrow H_2O + •OH + O_2$

$•OH + H_2O_2 \rightarrow H_2O + •OOH$

### ii. Scavenger test and Re-usability

The scavenger test is a suitable method to estimate the amount of radicals (hydroxyl, superoxide, and holes) that play a major role in the photocatalytic degradation of organic dye. In this test, three different types of chemicals, such as p-benzoquinone (BQ), ammonium oxalate (AO), and isopropanol (IPA), were employed to trap superoxide radicals, hole radicals, and hydroxyl radicals, respectively. The amount of different chemicals taken is the same as in our previous report [55]. It can be observed that the degradation percentage is lowest in the presence of IPA as hydroxyl radical scavengers [Figure 5(g)]. Therefore, the presence of IPA reduces hydroxyl radical concentration, drastically reducing the degradation process. Hence, this particular radical seems to be a determining factor in the degradation process.

Additionally, three cycles of the degradation process were run to ensure the stability and reusability of the nps for the MB dyes [Figure 5(i)]. The samples were collected after each cycle by centrifugation. After they were washed with ethanol and distilled water, they were dried in the microwave oven. Then, the samples were used again for photocatalytic dye degradation. It is seen that the efficiency of the prepared samples nominally decreases for the subsequent consecutive cycles. This may be due to the loss of nanomaterials during the filtering and washing process.

### b. OER activity

An aqueous solution of 1 M KOH was used to perform electrochemical studies of the as-synthesized sample for OER. The performance of all the samples for OER in an alkaline medium of 1 M KOH was explored for the examinations of LSV studies. In 1 M KOH, the LSV curve of all the samples clearly shows that the overpotential required to drive the current density of 10 mA $cm^{-2}$ is only 338 mV for the CF1 sample, which is much lower than that of other electrocatalysts C0 (415 mV), CF2 (367 mV), and the bench-mark $IrO_2$ (352 mV) [Figure 7(a)] [83]. The kinetic analysis used the Tafel equation ($\eta$ = blogj + a, where $\eta$, j, b, and a stand for the overpotential, current density, slope, and intercept, respectively. Tafel values were determined using the linear portion of the slope, which were 69, 103, and 117 mV $dec^{-1}$ for CF1, CF2, and C0, respectively, for OER [Figure 7(b)]. EIS was applied at the initial potential in the range of 100 kHz to 0.1 Hz. The EIS graph displays the high- and low-frequency ranges. In the high-frequency range, electrocatalysts exhibit short semicircles for OER, as shown in Figure 7c, to prove the small charge transfer. The semicircles show lower values of charge transfer resistance ($R_{ct}$) for CF1 (68.15 Ω) than those of other counterparts, C0 (170.01 Ω), and CF2 (74.75 Ω), signifying the fast kinetic process of CF1 catalyst occurred for the OER activity. In the low-frequency range, another semicircle was observed for all the samples, which could be due to the charge transfer between the diffused electrolyte and the bulk of the oxide materials [84], [85], [86]. The long-term stability of the CF1 electrocatalyst toward OER activity was also tested under the same experimental condition for 2500 continuous LSV cycles [Figure 7(d)]. We found a 7 mV increase in the overpotential at a current density of 10 mA cm$^{-2}$, which proved the long-term superior durability of the CF1 catalyst toward the OER activity in an alkaline medium. Furthermore, we have also studied the stability of the CF1 by chronopotentiometry method at a current density of 10 mA cm$^{-2}$ for 10 h [Insets of Figure 7(d)]. The result shows very high durability with negligible potential addition due to the resistance to corrosion behaviors of the CF1 catalysts during the electrocatalytic performance.

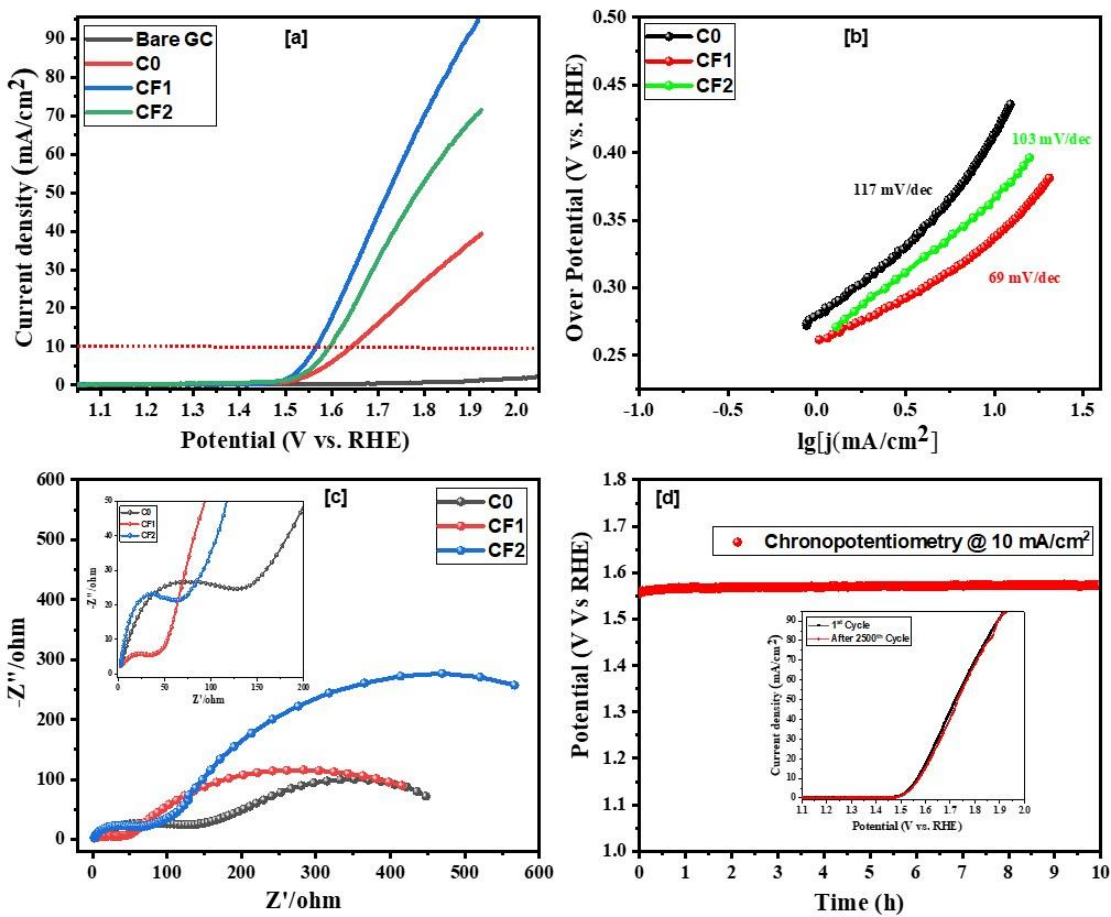

*Figure 7: represents (a) Polarization curves (LSV) plot, (b) corresponding Tafel plot (c) Impedance spectroscopy, where the insets show the high-frequency range semicircles of C0, CF1, and CF2 catalysts for OER and (d) chronopotentiometry study of CF1 for OER in 1.0 M KOH as a function of current density, where the insets show the Polarization curves of CF1 after 1st and 2500th cycles of continuous operation.*

The ECSA of the samples was determined by the $C_{dl}$ method, where the slope of the fitted curve was measured as ECSA. Short CV measurements were taken within non-Faradaic regions at sweep speeds of 20, 40, 60, 80, and 100 mVs$^{-1}$ within a potential window of 0.2 – 0.3 V (vs Hg/HgO) to calculate $C_{dl}$ [Figure 8(a, b, c)]. The ECSA value of CF1 is 76 cm$^2$, whereas for C0, it is 13.5 cm$^2$, and for CF2, it is 49 cm$^2$ [Figure 8(d)]. The higher ECSA value of CF1 may be because of the optimal concentration of Fe on the CuO surface, which acts as an active site for OER. Hence, the higher activity of CF1 towards OER as compared to C0 and CF2 can be justified by ECSA.

Mass activity (MA) and Specific activity (SA) were also determined for all the samples [83]. SA reveals the intrinsic activity of a catalyst [87], [88], [89], [90]. Considering the mass loading (m) on the GC to be 0.19 mg/cm$^2$, we have calculated the mass activity of the catalysts, where the CF1 exhibited higher mass activity at an overpotential of 338 mV (52.63 A.g$^{-1}$) compared to other electrocatalysts C0 (19.26 A.g$^{-1}$), and CF2 (30.57 A.g$^{-1}$). MA was calculated using the following equation: MA (A/g) = J (A/cm$^2$)/m (g/cm$^2$), where J is the current density at a particular overpotential, here 338, and m is the mass loading on the electrode. Similarly, SA is another important parameter normalized to the BET surface area of the catalysts [91]. SA was calculated using the following equation: SA (mA/ cm$^2$) = J (mA/cm$^2$)/10* $m$ $(g/cm^2)$ S$_{BET}$ (m$^2$/g), where S$_{BET}$ is the BET surface area obtained using N$_2$ adsorption-desorption data [92]. Again, CF1 exhibited higher mass activity at the same overpotential of 338 mV (0.82 mA/cm$^2$) compared to CF2 (0.63 mA/cm$^2$); however, it is less compared to C0 (1.29 mA/cm$^2$).

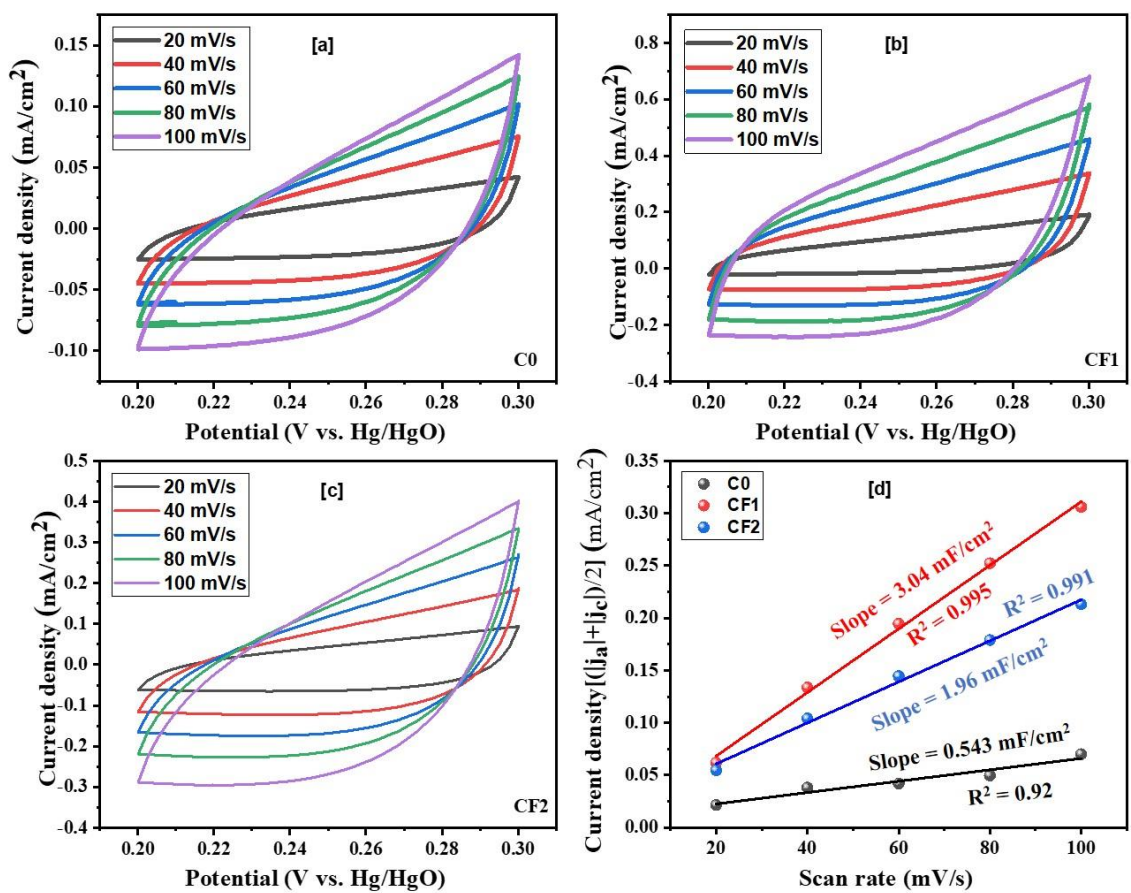

*Figure 8: (a, b, c) Cyclic voltammetry curves of C0, CF1, and CF2, respectively, and (d) corresponding mean plot of $J_a$ and $J_c$ against scan rate for the determination of double layer capacitance ($C_{dl}$) of the catalysts.*

5. Conclusion

In summary, we report the synthesis, characterization, and application of Iron-doped copper oxide (CuO) as low-cost catalysts for OER and wastewater treatments. The OER occurs at about 1.49 V versus the RHE ($\eta$ = 260 mV) with a Tafel slope of 69 mV dec$^{-1}$ in a 1 M KOH solution. The overpotential ($\eta_{10}$) of 338 mV at 10 mA cm$^{-2}$ is among the lowest compared with other copper-based materials. The catalyst can deliver a stable current density of >10 mA cm$^{-2}$ for more than 10 hours. The as-synthesized catalysts are also utilized for Fenton-like photo-degradation under low-power visible household LED lights toward the most commonly industrially used simulated Methylene blue dye wastewater. Almost complete degradation of the MB dye was achieved within 50 minutes of visible light irradiation with a first-order reaction with a rate constant of 0.0973 min$^{-1}$. We hope this dual functionality feature can open new pathways as a non-noble, highly efficient, and robust catalyst for OER and wastewater treatments.

**Acknowledgment:**

The authors want to acknowledge the Department of Science and Technology (DST), Govt of India, for providing the funds (DST/TDT/AMT/2017/200). SCB would like to thank DST INSPIRE for Providing fellowships (IF190617). The authors would like to acknowledge the Sophisticated Instrument Centre (SIC) facilities for providing the BET surface area analyzer facility at IIT Indore. The authors also acknowledge the DST, Govt. of India, New Delhi, India, for providing a FIST instrumentation fund to the discipline of Physics, IIT Indore, to purchase a Raman Spectrometer (Grant Number SR/FST/PSI-225/2016). The Authors would also like to thank Dr. Mrigendra Dubey, Associate Professor in MEMS, IIT Indore, for FTIR spectroscopy measurements and Mukesh Yadav for technical support. The Authors are deeply thankful to Dr. Rupesh S. Devan, Associate professor in the department of MEMS, IIT Indore, for UV-vis measurements. The Authors want to acknowledge Dr. Jaydeep Bhattacharya, School of Biotechnology, Jawaharlal Nehru University, New Delhi, India, for DLS measurements.